\documentclass[aps,prl,twocolumn,superscriptaddress]{revtex4}
\usepackage{amsmath,amsfonts,amssymb}
\usepackage{graphicx,subfigure}

\begin{document}

\title{Wave-particle duality revisited}

\author{Uwe Schilling}
\affiliation{Institut f\"{u}r Optik, Information und Photonik and Erlangen Graduate School in Advanced Optical Technologies (SAOT), Universit\"{a}t Erlangen-N\"{u}rnberg, 91058 Erlangen, Germany}

\author{Joachim von Zanthier}
\affiliation{Institut f\"{u}r Optik, Information und Photonik and Erlangen Graduate School in Advanced Optical Technologies (SAOT), Universit\"{a}t Erlangen-N\"{u}rnberg, 91058 Erlangen, Germany}

\date{\today}

\begin{abstract}
We investigate wave-particle duality in a symmetric two-way interferometer with a which-way detector. We find that it is important to state wether the interfering object or the which-way detector is read out first. In case that the interfering object is read out first, we discover that it is possible to use the information about its state to increase the distinguishability to a value which violates the widely accepted bounds set by an inequality introduced by Jaeger \emph{et al.}~\cite{Jaeger:1995} and Englert~\cite{Englert:1996}.
\end{abstract}


\maketitle

The concept of wave-particle duality lies at the heart of quantum mechanics and has been subject to a long and ongoing debate~\cite{Bohr:1949,Wootters:1979,Greenberger:1988,Zeilinger:1986,Mandel:1991,Jaeger:1995,Englert:1996,Kolar:2007,Afshar:2007,Sen:2009}. It describes the trade-off between the particle properties and the wave properties of a quantum mechanical object (a quanton) in a two-way interferometer, the former commonly related to the information about the path of the object, the latter to the ability to interfere on a detector behind the interferometer, manifesting itself in the visibility of an interference pattern. The most widely accepted description of duality has been developed independently by Jaeger \emph{et al.}~\cite{Jaeger:1995} and Englert~\cite{Englert:1996}. It is expressed by the formula
\begin{align}
\mathcal D^2 + \mathcal V^2 \leq 1
\label{eq_duality}
\end{align}
describing how the which-way (WW) information as measured by a which-way detector (WWD) and quantified by the distinguishability $\mathcal D$ between the two paths limits the visibility $\mathcal V$ of the interference pattern and vice versa. This inequality has been confirmed in numerous experiments~\cite{Duerr:1998,Duerr:1998b,Buks:1998,Schwindt:1999,Pryde:2004,Peng:2005,Jacques:2008,Barbieri:2009} and theoretical investigations~\cite{Bjoerk:1998,Abranyos:1999,Englert:2000,Miniatura:2007,Erez:2009}. However, often both quantities are measured only in seperate runs of the experiment. In contrast, the very idea of duality implies that both quantities are measured simultaneously, i.e. for a complete run of the experiment, both the WWD and the interfering quanton should be read out. The question then appears which of the two quantities should be measured first, since, as already pointed out by Wootters and Zurek, the quanton and the WWD have ``become nonseparable parts of a single quantum-mechanical system'' and ``this forces us to consider the effect of our measurement''. In this paper, we demonstrate that in general it is crucial to state wether the WWD or the state of the quanton on the detector behind the interferometer is read out first, as we arrive at different values for the distinguishability $\mathcal D$ in these two scenarios.

\begin{figure}[ht]
\centering
\subfigure[ ]{\includegraphics[scale=0.35]{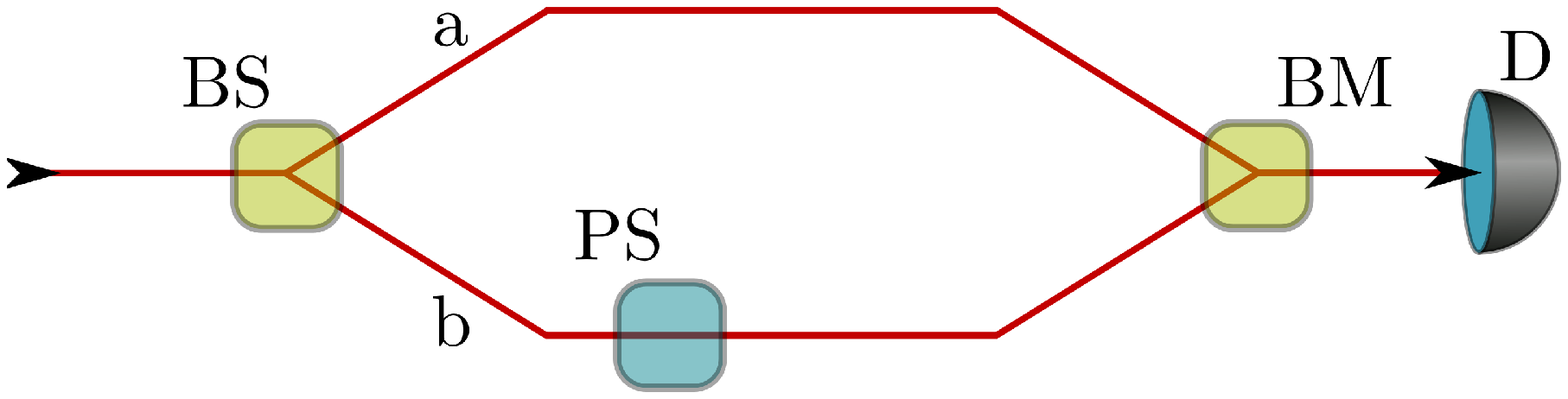}\label{fig_int_simple}}
\subfigure[ ]{\includegraphics[scale=0.35]{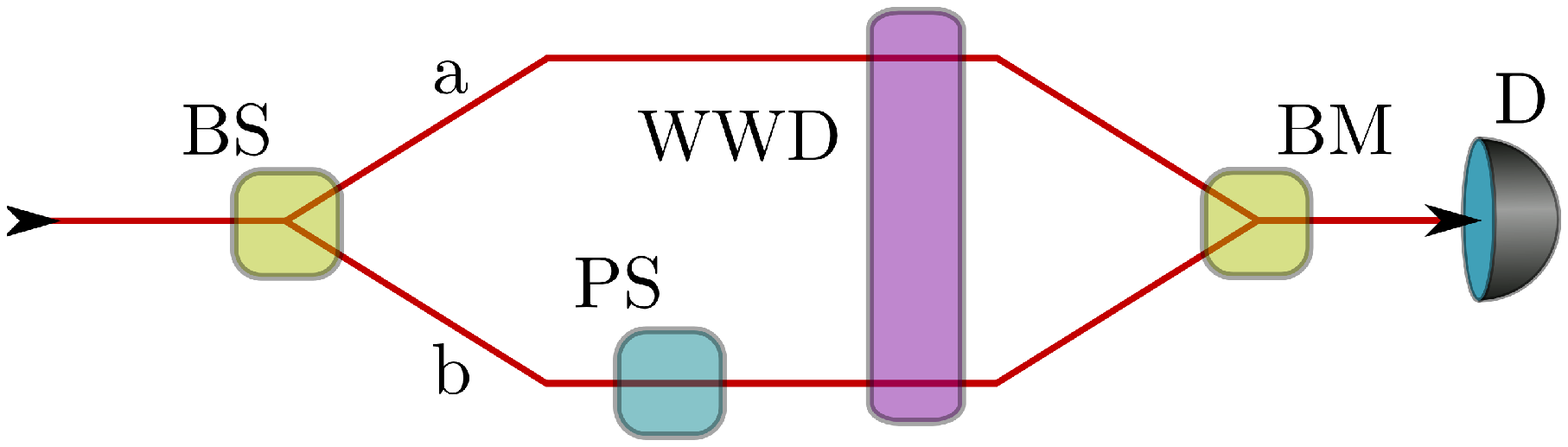}\label{fig_int_WWD}}
\subfigure[ ]{\includegraphics[scale=0.35]{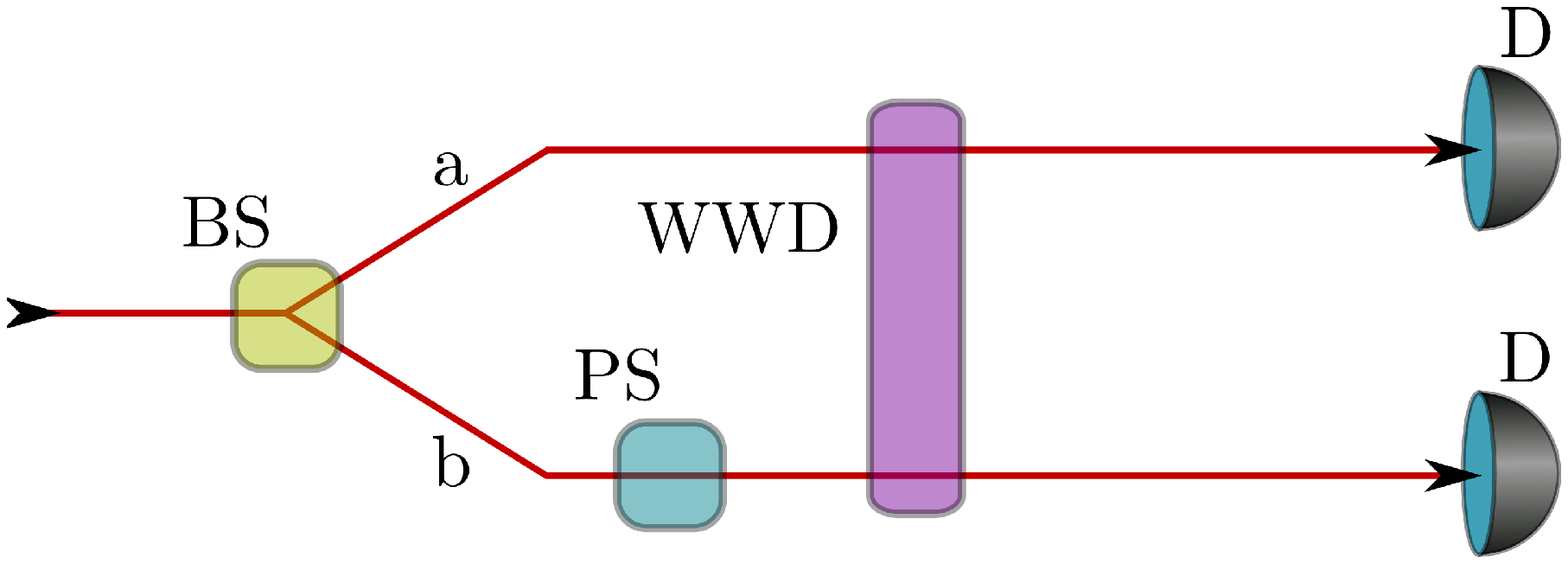}\label{fig_int_path_measurement}}
\caption{(a) Setup of a very general two-way interferometer where the incoming quanton encounters a beam splitter (BS) and propagates along two different paths $a$ and $b$ with a relative phase shift (PS). Subsequently, the two beams are recombined at the beam merger (BM) and measured at the detector behind the interferometer (D). (b) Same setup as in (a), but including a WWD, which interacts on both paths with the quanton. (c) In order to verify the WW information gained from the WWD, we need to measure the occupation of the two modes $a$ and $b$ instead of the interference pattern.}
\label{fig_setup}
\end{figure}
In general, a symmetric two-way interferometer can be described as follows: the quanton in the initial state $|\psi_a\rangle$ enters the interferometer and encounters a 50:50 beam splitter (BS) where its wave function is split equally into two orthogonal parts $|\psi_a\rangle$ and $|\psi_b\rangle$. The splitting may refer to an actual spatial seperation of the wave function as well as to different internal states of the quanton. Following Englert's notation~\cite{Englert:1996}, we describe the transformation at the beam splitter by a Hadamard gate $H$ acting on the incoming state $|\psi_a\rangle$
\begin{align}
|\psi_a\rangle \stackrel{\text{BS}}{\rightarrow} H|\psi_a\rangle = |\psi_a\rangle + |\psi_b\rangle.
\end{align}
(Throughout the paper we will neglect normalization constants.) In the central stage of the interferometer, the two parts of the wave function experience a certain relative phase shift $\delta$
\begin{align}
|\psi_a\rangle + |\psi_b\rangle \stackrel{\text{PS}}{\rightarrow} |\psi_a\rangle + e^{i \delta}|\psi_b\rangle;
\label{eq_after_PS}
\end{align}
the two paths are then recombined at the beam merger (BM), an action which is again described by a Hadamard gate
\begin{align}
|\psi_a\rangle + e^{i \delta}|\psi_b\rangle \stackrel{\text{BM}}{\rightarrow} (1+e^{i\delta})|\psi_a\rangle + (1-e^{i\delta})|\psi_b\rangle,
\end{align}
and finally, the interference pattern is revealed at the detector in the relative frequency $P_{a}(\delta)$ of finding state $|\psi_a\rangle$ (and of course also of its complement $|\psi_b\rangle$) on varying the interferometric phase $\delta$ (cf. Fig. \ref{fig_int_simple}).

In order to gain information about the path of the quanton in a symmetric interferometer, one must either use an unbalanced beam merger or introduce a WWD into the arms of the interferometer~\cite{Jacques:2008}. Here, we discuss the latter case (cf. Fig. \ref{fig_int_WWD}) where we assume that the WWD is a device which records the path of the quanton without changing the state of the quanton. Quantum mechanically, this WWD can be described as an object initially in state $|\chi^{(i)}\rangle$ which upon interaction with the quanton in the state $|\psi_a\rangle$ ($|\psi_b\rangle$) is transfered into state $|\chi_a\rangle$ ($|\chi_b\rangle$). Hereby the two final states $|\chi_a\rangle$ and $|\chi_b\rangle$ of the WWD are, in contrast to the states $|\psi_a \rangle$ and $|\psi_b \rangle$ of the quanton, not necessarily orthogonal. This allows for a partial measurement with a principally suboptimal WWD.

The action of the WWD commutes with the one of the phase shifter and consequently their order is of no importance; after both elements, the state of the combined system of WWD and quanton is given by
\begin{align}
(|\psi_a\rangle + e^{i\delta}|\psi_b\rangle) \otimes |\chi^{(i)}\rangle \stackrel{\text{WWD}}{\rightarrow} |\psi_a\rangle |\chi_a\rangle + e^{i \delta}|\psi_b\rangle |\chi_b\rangle.
\label{eq_WW_after_PS}
\end{align}
Finally, the beam merger acts as a second Hadamard gate on the state of the quanton, transforming the combined system into its final state $|\Psi^{(f)}\rangle$:
\begin{multline}
|\psi_a\rangle |\chi_a \rangle + e^{i \delta}|\psi_b\rangle |\chi_b \rangle \stackrel{\text{BM}}{\rightarrow}\\
|\psi_a\rangle |\chi_a \rangle + |\psi_b\rangle |\chi_a \rangle + e^{i \delta} |\psi_a\rangle |\chi_b \rangle - e^{i \delta} |\psi_b\rangle |\chi_b \rangle = |\Psi^{(f)}\rangle.
\label{eq_WW_after_BM}
\end{multline}
The visibility of the interference pattern $P_{a}(\delta)$ is now given by
\begin{align}
\mathcal V := \frac{P_{a,\text{max}}-P_{a,\text{min}}}{P_{a,\text{max}}+P_{a,\text{min}}}= |\langle \chi_a | \chi_b\rangle |
\label{eq_visibility}
\end{align}
where $P_{a,\text{max}}$ ($P_{a,\text{min}}$) denotes the maximal (minimal) value of $P_{a}(\delta)$. To make the further analysis more transparent, we write the state of the WWD in a basis that reveals explicitly which parts of the wave function of $|\chi_a\rangle$ and $|\chi_b\rangle$ overlap and which do not:
\begin{align}
\begin{array}{l}
|\chi_a\rangle = \alpha_a |0 \rangle + \beta_a |+\rangle \hspace{2em} \text{and}\\
|\chi_b\rangle = \alpha_b |0 \rangle + \beta_b |-\rangle,
\end{array}
\end{align}
where the three states $|0\rangle$, $|+\rangle$, and $|-\rangle$ are pairwise orthogonal. We will call this basis the natural basis. Using Eq.~(\ref{eq_visibility}), the visibility of the interference pattern can now be cast into the form
\begin{align}
\mathcal V = |\alpha_a^* \alpha_b|.
\label{eq_visibility2}
\end{align}
This result is not unexpected, since less overlap between $|\chi_a\rangle$ and $|\chi_b\rangle$ implies a better distinguishability between the two paths and vice versa. For example, if $|\chi_a\rangle$ and $|\chi_b\rangle$ are orthogonal, reading out the WWD in this basis will give full information about the path of the quanton in the central stage of the interferometer while no interference pattern will appear. However, in general, it is necessary to first specify an observable with an eigenbasis $W = \{|\chi_k\rangle, k=1,\ldots,n\}$ in which to read out the WWD. The subsequent readout process of the observable provides, in most cases incomplete, information about the path of the quanton. One can quantify the amount of WW information in such a measurement by the chance $L_j$ of guessing the way correctly under the condition that the WWD is found in a certain basis state $|\chi_j\rangle$ ($j\in \{1,\ldots,n\}$)~\cite{Englert:1996}. $L_j$ is determined by the maximum of the two probabilities $p_a$ and $p_b$ of finding the quanton in either state,
\begin{align}
L_j = \max (p_a,p_b) = \frac{\max\left( \left|\langle\chi_a | \chi_j \rangle\right|^2,\left|\langle \chi_b | \chi_j \rangle\right|^2 \right)}{\left|\langle \chi_a | \chi_j \rangle\right|^2 + \left|\langle \chi_b | \chi_j \rangle\right|^2}.
\label{eq_q_i}
\end{align}
The likelihood $\mathcal L$ of guessing the state of the quanton correctly in an arbitrary measurement of the WWD is then given by weighting all $L_j$ with the probability of finding the WWD in the corresponding state $|\chi_j\rangle$:
\begin{align}
\mathcal L = \sum_{j=1}^n L_j \langle \chi_j|\rho_D|\chi_j \rangle,
\label{eq_destilled_likelihood}
\end{align}
where the density matrix $\rho_D$ denotes the final state of the WWD. Experimentally, $\mathcal L$ may be verified by a setup as shown in Fig. \ref{fig_int_path_measurement}, where the occupation probabilities of the two paths are determined from the detector readout results. $\mathcal L$ is a measure for the amount of WW information that can be \emph{extracted} from the WWD in this basis: If $\mathcal L = 0.5$, we have to make a random guess about the path of the quanton, whereas for $\mathcal L = 1$ we know the way with certainty. Maximizing $\mathcal L$ with respect to $W$ leads to the optimal likelihood $\mathcal L_{\text{opt}}$ which quantifies the amount of WW information that is principally \emph{available}. In order to have a measure which has a value of 0 (1) for no (full) WW information, the distinguishability $\mathcal D$ is introduced, which rescales $\mathcal L_{\text{opt}}$ to the interval from 0 to 1~\cite{Englert:1996}:
\begin{align}
\mathcal D = 2 \mathcal L_{\text{opt}} - 1 = 2 \sum_{j=1}^n L_j \langle \chi_j^{\text{opt}}|\rho_D|\chi_j^{\text{opt}} \rangle - 1.
\label{eq_distinguishability}
\end{align}

As can be seen from this equation, $\mathcal D$ depends on the final state of the WWD. However, the final state of the WWD is not well defined without stating wether the quanton has already been measured. This is easily understood by realizing that Eq.~(\ref{eq_WW_after_PS}) describes an entangled state whenever $|\langle \chi_a | \chi_b \rangle|^2 \neq 1$ and in such an entangled state the measurement of one part of the system changes the state of the other part. Thus, we need to specify wether one reads out the WWD right away and then lets the quanton propagate through the beam merger to observe the interference fringes, or wether one first has the quanton measured at the detector behind the interferometer and then reads out the WWD afterwards.

\subsection{Reading out the WWD first}
We start with the scenario in which the WWD is read out first. As the quanton has not been measured yet, the state of the WWD is derived from Eq. (\ref{eq_WW_after_BM}) by tracing over the degrees of freedom of the quanton:
\begin{align}
\rho_D = \text{tr}_Q\{|\Psi^{(f)}\rangle \langle \Psi^{(f)}|\} = |\chi_a\rangle\langle\chi_a| + |\chi_b\rangle\langle\chi_b|
\label{eq_final_detector_state_first}
\end{align}
In that case, it was shown in Ref. \cite{Englert:1996} that in order to access all available WW information, the basis for reading out the WWD has to be chosen as the eigenvectors of the matrix $\rho = |\chi_a\rangle\langle\chi_a| - |\chi_b\rangle\langle \chi_b|$ and that $\mathcal D$ is given by
\begin{align}
\mathcal D = \frac{1}{2} \text{tr}\{\left| \rho \right|\} = \sqrt{1-|\langle\chi_a|\chi_b\rangle|^2}.
\label{eq_englert_D}
\end{align}
After having read out the WWD, one may observe the interference of the quanton after the beam merger where one finds an interference pattern which has a visibility $\mathcal V$ given by Eq. (\ref{eq_visibility}). Obviously, Eq.~(\ref{eq_duality}) is in this case always optimally fulfilled.

\subsection{Reading out the quanton first}
If the quanton is read out first, one encounters a different situation: Reading out the state of the quanton leads to a projection of the state $|\Psi^{(f)}\rangle$ in Eq.~(\ref{eq_WW_after_BM}) where the projected state depends on wether state $|\psi_a\rangle$ or state $|\psi_b\rangle$ is measured. In this case, after the measurement of the quanton, the state of the WWD is described by
\begin{align}
|\chi_a\rangle + \sigma e^{i \delta} |\chi_b\rangle
\label{eq_final_detector_state_second}
\end{align}
where $\sigma = 1$ ($\sigma = -1$) if the state of the quanton was measured to be $|\psi_a\rangle$ ($|\psi_b\rangle$). Note that an equally weighted statistical mixture of these two possible final states was found in Eq. (\ref{eq_final_detector_state_first}) for the case that there is no information about the state of the quanton, i.e., if one does not use the information provided by the readout of the quanton, one recovers the results from the previous section. However, the measurement has been performed, and reading out the result means that the state of the WWD is actually given by the pure state described by Eq.~(\ref{eq_final_detector_state_second}). Since this state is different from the one in Eq. (\ref{eq_final_detector_state_first}), we can try to find a basis which performs better than the one used for the derivation of Eq. (\ref{eq_englert_D}). In view of Eq.~(\ref{eq_final_detector_state_second}), this basis might depend on the result of the quanton readout and the interferometric phase $\delta$. Using the natural basis, Eq. (\ref{eq_final_detector_state_second}) transforms into
\begin{align}
|\chi^{(f)}\rangle = (\alpha_a + \sigma e^{i\delta} \alpha_b) |0\rangle + \beta_a |+\rangle + \sigma e^{i \delta} \beta_b |-\rangle.
\label{eq_final_detector_state_second_new}
\end{align}
We may now try to maximise Eq.~(\ref{eq_distinguishability}) for the distinguishability $\mathcal D$ with regard to the measurement basis where the state of the detector is now given by $\rho_D = |\chi^{(f)}\rangle\langle \chi^{(f)}|$. To elucidate this point, let us consider the case of a symmetric WWD where $|\alpha_a| = |\alpha_b|$; this can be realized, for example, by placing two identical and independent WWD devices in each arm of the interferometer. In this case, it is obvious from Eq.~(\ref{eq_final_detector_state_second_new}) that for certain values of the phase shift $\delta$ the contribution of the state $|0\rangle$ to the final state vanishes. On reading out the WWD in the natural basis one will thus always find either $|+\rangle$ or $|-\rangle$, both of which can be unambiguously identified with either path $|\psi_a\rangle$ or $|\psi_b\rangle$. Consequently, we have a distinguishability of 100\,\% while, according to Eq. (\ref{eq_visibility2}), the visibility of the interference pattern can take on principally any value between 0 and 100\,\%~\footnote{However, if the visibility is 100\,\%, one can show that there will be no signal at that value of $\delta$.}. For this specific value of $\delta$, we thus found a WWD basis which contains more WW information than the one used in Eq.~(\ref{eq_englert_D}). However, we stress that it is only possible to find such a basis because of the additional information we have about the state of the quanton after the measurement process and the corresponding projection of the WWD state.

\begin{figure}[t]
\centering
\subfigure{\includegraphics[scale=0.38]{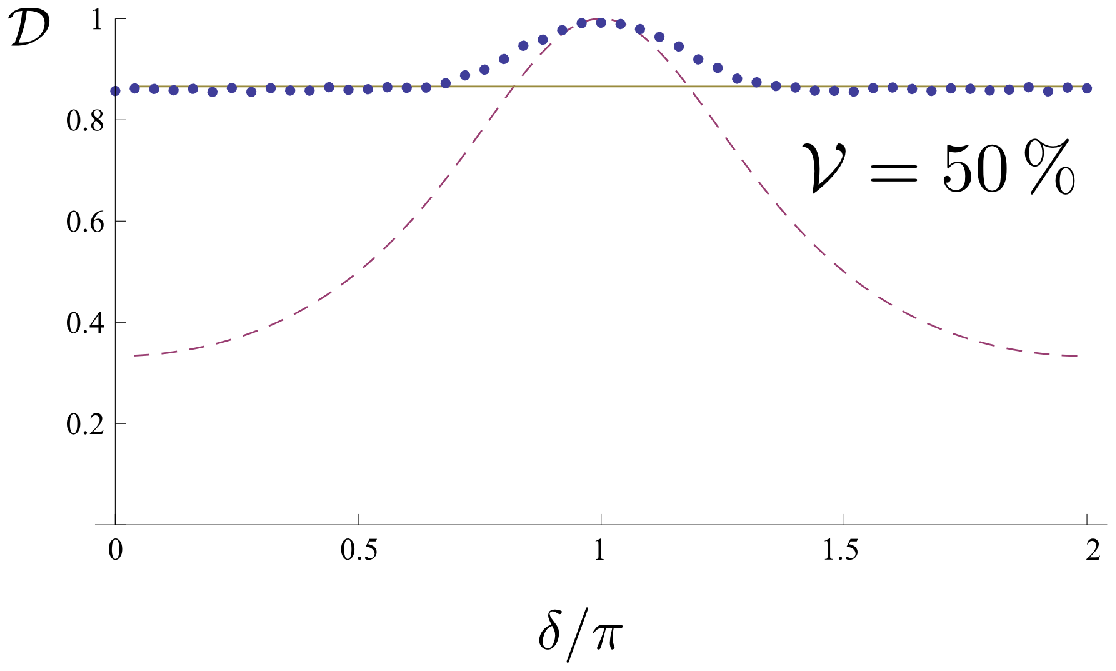}}
\subfigure{\includegraphics[scale=0.38]{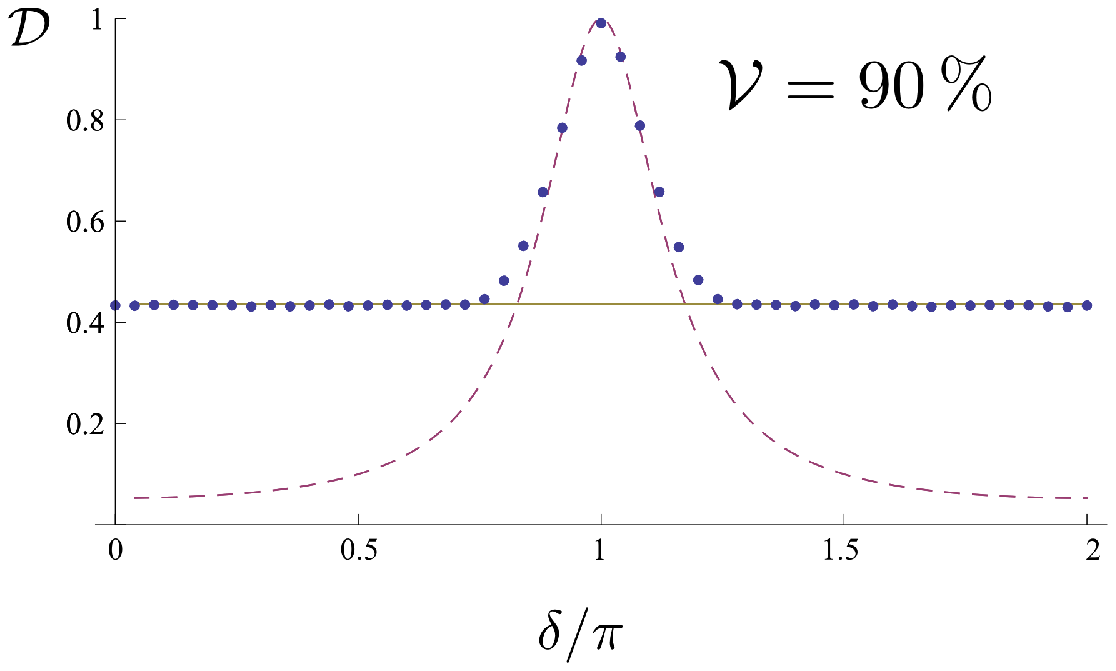}}
\subfigure{\includegraphics[scale=0.38]{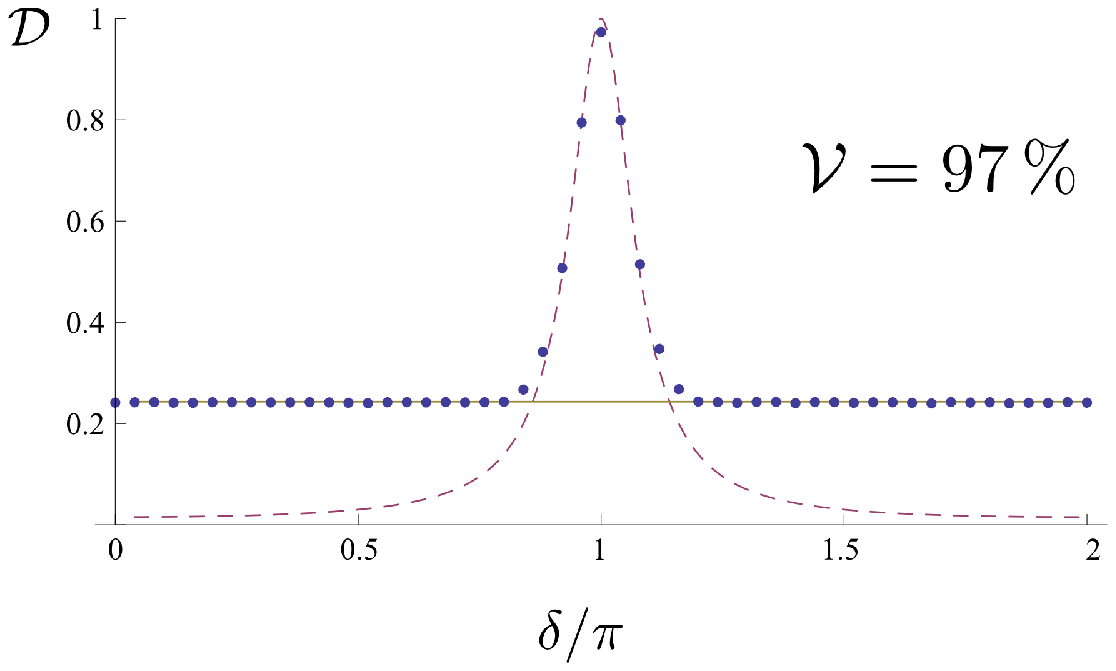}}
\caption{(color online). Numerical results for the value of $\mathcal D$ for a symmetric two-way interferometer with a symmetric WWD, if the state $|\psi_a\rangle$ is found (dots). Also shown are the values of $2 \mathcal L -1$ using Englert's basis (full black line) or the natural basis (dashed purple line). For the case that the quanton was found to be in state $|\psi_b\rangle$ the graphs are identical but shifted by $\pi$.}
\label{fig_final_result}
\end{figure}

For other values of $\delta$, the ideal basis is less obvious. To find $\mathcal L_{\text{opt}}$ and thus also $\mathcal D$ for any phase shift, we performed a Monte Carlo simulation in which $\mathcal L$ was calculated for 10.000 random bases for fifty different values of $\delta$ at a given value of $\mathcal V$. The results for $\mathcal V = 50\,\%$, $\mathcal V = 90\,\%$, and $\mathcal V = 97\,\%$ are shown in Fig. \ref{fig_final_result}.
As can be seen from the figure, the distinguishability has now become a quantity which depends on $\delta$ and $\mathcal V$. In addition, we can note that there is a range around the maximum of $\mathcal D$ in which the optimal basis is very close to the natural one and a region along the ``wings'' of the plot of $\mathcal D$ in which it is close to the one used in Eq.~(\ref{eq_duality}). Only in a small intervall between the two regions there is a deviation where a third basis should be used in order to obtain a value which is noticeably closer to $\mathcal D$. Unfortunately, we cannot present an analytical expression for that basis and the corresponding distinguishability. However, the central result is that for all phase shifts $\delta$, we find a distinguishability \emph{at least} as high as in the scenario which was used to derive Eq.~(\ref{eq_englert_D}) and for certain phase shifts, we are able to violate the limits set by Eq. (\ref{eq_duality}).  

The conclusions one can draw from these results are two-fold. First, by applying well-established procedures for the derivation of $\mathcal D$ and $\mathcal V$, we find that it does make a difference wether the quanton is read out before the WWD or vice versa. This is in contrast to what is sometimes stated for specific setups (see e.g. \cite{Wootters:1979,Herzog:1995,Miniatura:2007}). Second, one can raise the valid questions wether the procedures introduced in \cite{Englert:1996} are applicable in the case that the quanton is measured first and wether in that case we still do measure WW information. The latter question arises because the scenario exploits an effect which occurs only because of the equally weighted interference of two paths in order to increase the WW information and then claim a posteriori that the quanton has been more likely to have taken one particular path. If one answers this question to the positive, then one must conclude that this scenario does allow for an amount of WW information which violates the limit set by Eq.~(\ref{eq_duality}). If, on the other hand, one answers this question to the negative, then there is at least the important conclusion to be drawn, namely that it is of utmost importance to pay attention to when the WWD is read out, there is only a meaningful result for an experiment on wave-particle duality if the WWD is read out before the quanton.

In summary, we have shown that in a two-way interferometer with a WW detector the distinguishability $\mathcal D$ of  the two paths depends on the order in which the measurement of the states of the quanton and the WWD is performed. Reading out the WWD first leads to the well-known duality relation Eq.~(\ref{eq_duality}) as introduced in~\cite{Jaeger:1995,Englert:1996}. However, reading out the quanton first allows for a distinguishability exceeding the limit set by Eq. (\ref{eq_duality}). This increase is enabled by using the additional information provided by the readout of the quanton state as an input for the choice of the optimal basis in which to read out the detector. We leave open for discussion if the information one gains from reading out the WWD can still be considered WW information. But wether one answers this question positively or negatively, in both cases one has to conclude that the order of the measurement has decisive consequences to the amount of WW information that is principally available in a two-way interferometer.

The authors gratefully acknowledge funding of the Erlangen Graduate School in Advanced Optical Technologies (SAOT) by the German Research Foundation (DFG) in the framework of the German excellence initiative. US thanks the Elite Network of Bavaria for financial support.
\bibliography{bibliography}
\end{document}